%% file: bjr_dpf2019_proceedings.tex
\documentclass[11pt]{article}
\usepackage[margin=1in]{geometry}
\usepackage{graphicx}
\usepackage{units}
\usepackage{amsmath}
\usepackage{subcaption}

\usepackage[
    type={CC},
    modifier={by},
    version={4.0},
]{doclicense}

\makeatletter
\def\blfootnote{\gdef\@thefnmark{}\@footnotetext}
\makeatother

\usepackage[backend=biber]{biblatex}
\input econfmacros.tex 

\def\Title#1{\begin{center} {\Large {\bf #1} } \end{center}}
\def\Author#1{\begin{center} {\normalsize {\sc #1} } \end{center}}
\def\Institution#1{\begin{center} {\normalsize {\it #1} } \end{center}}
\def\Abstract#1{\noindent {\normalsize {\bf Abstract:} {\normalfont #1}}}
\def\Conference{\vspace{4mm}\begin{raggedright} {\normalsize {\it Talk presented at the 2019 Meeting of the Division of Particles and Fields of the American Physical Society (DPF2019), July 29--August 2, 2019, Northeastern University, Boston, C1907293.} } \end{raggedright}\vspace{4mm}}

\begin{document}

%
%

\Title{Verification of Readout Electronics in the ATLAS ITk Strips Detector}

\Author{Benjamin John Rosser}

\Institution{Department of Physics and Astronomy\\ University of Pennsylvania}

\Abstract{Particle physics detectors increasingly make use of custom FPGA firmware
and application-specific integrated circuits (ASICs) for data readout and triggering.
As these designs become more complex, it is important to ensure that they are simulated
under realistic operating conditions before beginning fabrication. One tool
to assist with the development of such designs is cocotb, an open source digital logic
verification framework. Using cocotb, verification can be done at high level using
the Python programming language, allowing sophisticated data flow simulations to be
conducted and issues to be identified early in the design phase. Cocotb was used
successfully in the development of a testbench for several custom ASICs for
the ATLAS ITk Strips detector, which found and resolved many problems during
the development of the chips.}

\Conference \blfootnote{Copyright owned by the authors. \doclicenseText}

%
%

\section{Introduction}

Work is currently underway in preparation for the High Luminosity upgrade to
the Large Hadron Collider (HL-LHC), scheduled to begin running in 2026. This
upgrade project will increase the number of proton-proton collisions by a
factor of 3.5. By the end of the HL-LHC's run in the late 2030s, an order of
magnitude more data will have been collected.
In order to run at this higher collision rate, the ATLAS detector \cite{PERF-2007-01}
requires a number of upgrades to enable faster data processing and increased
radiation tolerance. Of the projects underway, the most significant is the
complete replacement of the ATLAS Inner Detector, which sits closest to the
beam and is used to track electrically charged particles as they travel
through the detector.

Currently under development is the brand new, all-silicon ATLAS Inner
Tracker (ITk), comprised of pixel \cite{Collaboration:2285585} and strip \cite{Collaboration:2257755} layers.
Both pixel and strip layers will use custom application-specific integrated
circuits (ASICs) as front-end readout. These ASICs will be responsible for
digitizing hits and reading out events in response to trigger commands. Their
designs are quite complex: they will need to read out events at a $\unit[1]{MHz}$
readout rate and also survive total radiation doses up to $\unit[1.7]{Grad}$
(in the pixels) and $\unit[50]{Mrad}$ (in the strips). Due to the complexity
and high cost involved in producing ASICs, careful testing and validation
of the design before manufacturing the chips is critical. Section \ref{sec:strips}
will discuss the readout chips in the strips detector in more detail,
and then Section \ref{sec:cocotb} will discuss the techniques used to
verify that design.

\section{ITk Strips Detector}
\label{sec:strips}

\begin{figure}[!htb]
    \centering
    \includegraphics[width=3.5in]{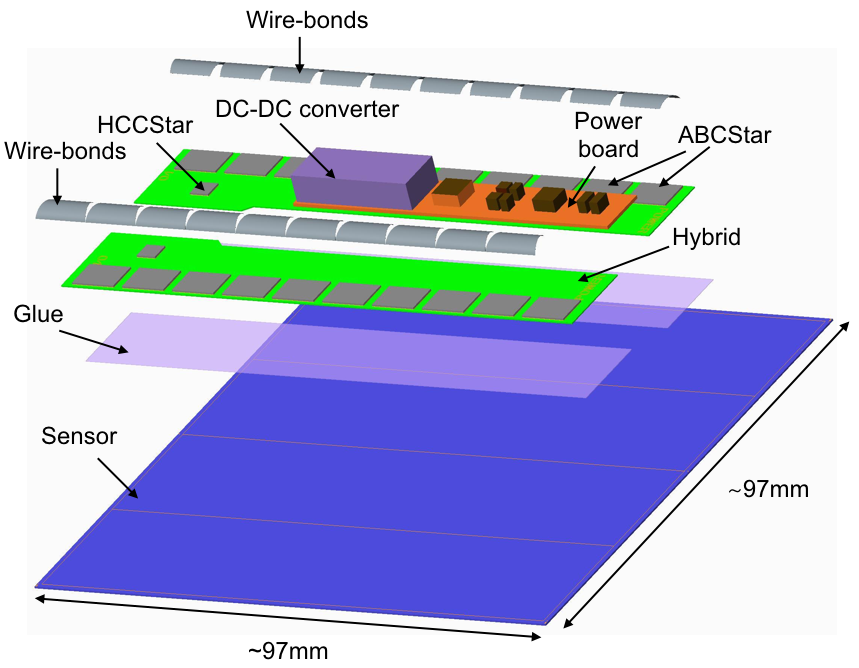}
    \caption{Diagram of an ITk Strips module from one of the inner
    barrel layers. Both front-end readout chips sit on an electrical
    hybrid (green) above the silicon sensors (blue). Depending on the
    exact geometry, modules can have one or two hybrids.}
    \label{fig:module}
\end{figure}

The strips detector is broken up into modules of varying geometries. Figure
\ref{fig:module} shows an example module from the barrel region.

Each module contains two different ASICs used in the readout of data.
ATLAS Binary Chips (ABCStars) digitize hits from the silicon strip sensors
and cluster the data. A hybrid contains a group of 6 to 11 ABCStars and a
Hybrid Controller Chip (HCCStar), which receives trigger commands
and dispatches them to each attached ABCStar. These ASICs form a
star network, hence their names, as the HCCStar receives clusters
back from the ABCStars in parallel. These asynchronous streams
need to be combined into data packets and then transmitted out of
the detector.

The HCCStar also supports a multi-level triggering mode. In this mode,
track information from up to 10\% of the detector would be used to make
the Level 1 (L1) triggering decision. The HCCStar thus has to support
two types of readout commands: regional readout requests (R3s), which
will have higher priority than the normal command, a L1 trigger.

\begin{figure}[!htb]
    \centering
    \includegraphics[width=3.5in]{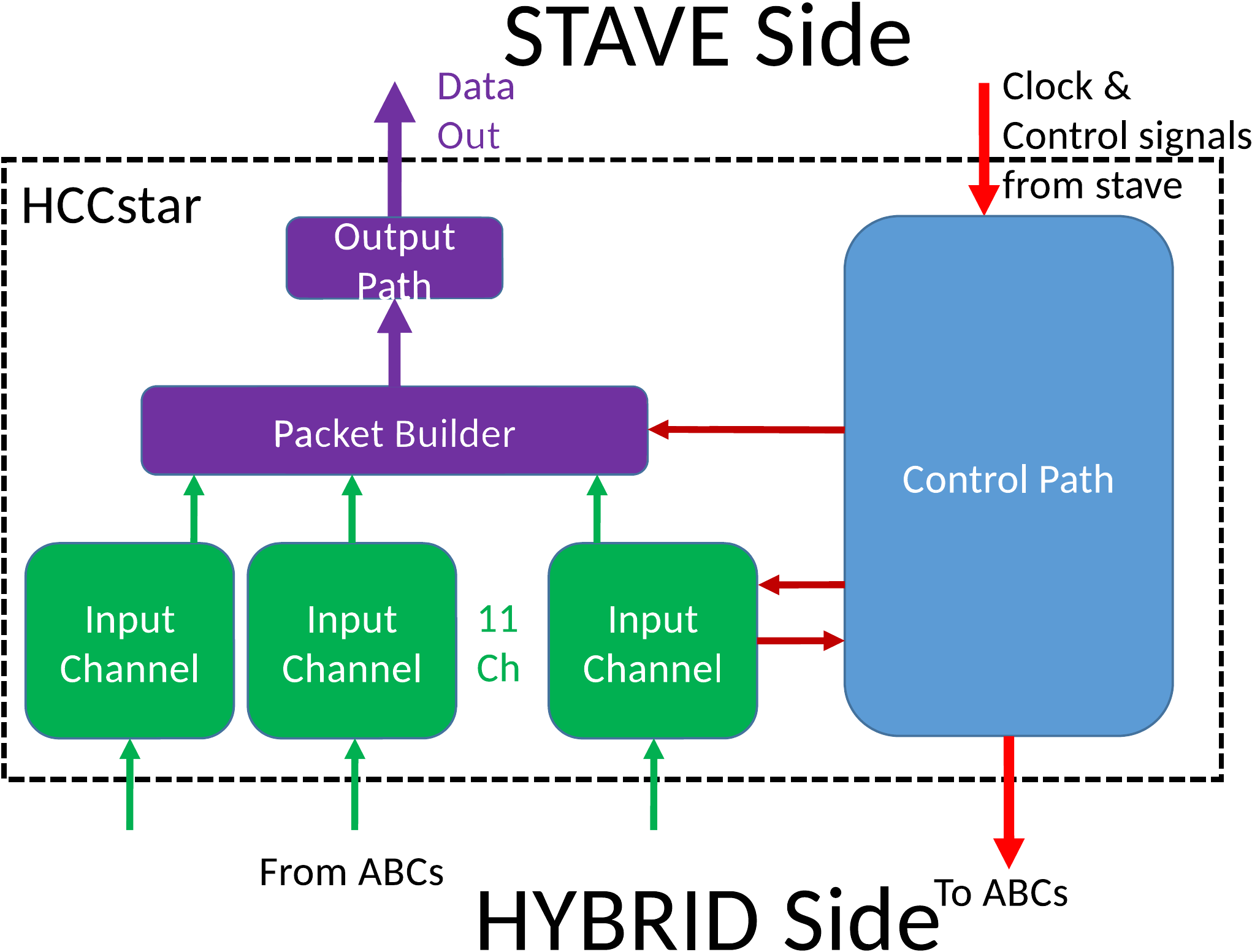}
    \caption{High-level logic diagram of the HCCStar, showing the flow of
    control signals from the ``stave side" to the ABCStars on the ``hybrid
    side", which then produce data packets. The packet builder (purple)
    is responsible for merging the ABCStar input streams into single
    output packets for every event, and is one of the most complex parts of the chip.}
    \label{fig:hccstar}
\end{figure}

A high-level logic diagram of the HCCStar is shown in Figure \ref{fig:hccstar}.
To support all these features, a large amount of digital logic is required.
The HCCStar's digital components were written in the hardware description
languages Verilog and VHDL. The codebase for the HCCStar engineering
prototype consisted of over 12,000 lines of Verilog and nearly 5,000 lines of
VHDL. Fabricating these ASICs costs about $\$350\text{k}$, which
makes it critical to simulate and verify the designs before submitting
them for production.

\section{Digital Logic Verification}
\label{sec:cocotb}

Digital verification is typically done by writing a simulation framework
(known as a ``testbench") using one of two approaches:

\begin{itemize}
\item Write a testbench by hand in a hardware description language. This
is ideal for simple designs and is relatively straightforward, but it
is difficult to scale up to complex designs.
\item Use a verification library as the starting point for the
testbench. Frameworks like the Universal Verification Methodology (UVM) \cite{7932212},
an industry-standard approach to digital logic verification, enable
very powerful testbenches to be created.
\end{itemize}

For the HCCStar, the first approach was rejected due to the large amount
of digital logic that required verification. The second approach was
considered, but ultimately rejected as well due to the high complexity
and learning curve of UVM.

Instead, the HCCStar verification was done using the open-source
package cocotb \cite{cocotb} (``coroutine cosimulation testbench"). Cocotb
makes it possible to write the testbench code in the Python programming
language, thus separating the problems of hardware design (which is still
done using a hardware description language) and hardware simulation (which
can now be done using a general-purpose programming language).

\begin{figure}[!htb]
    \centering
    \includegraphics[width=2.5in]{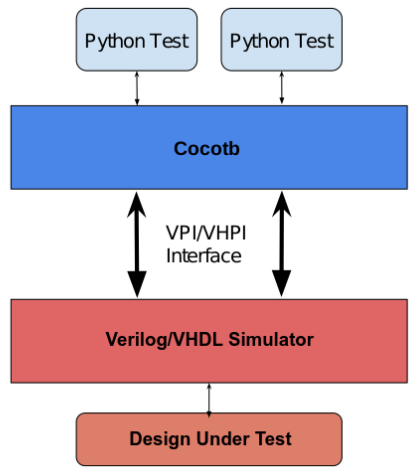}
    \caption{Architecture of a standard cocotb testbench. The design under test
    (DUT) is cosimulated using a standard Verilog or VHDL simulator (the HCCStar
    verification was done using the Cadence Incisive simulator) and controlled via
    Python.}
    \label{fig:cocotb}
\end{figure}

Figure \ref{fig:cocotb} shows the rough architecture of a cocotb testbench.
Communication between the Verilog/VHDL simulator (where the digital design
is running) and the Python test code is mediated by the cocotb library. When
Python code is executing, the simulator is paused: the testbench must
explicitly yield control back to the simulator to allow any changes to digital
signals to propagate. The testbench can then wait for a triggering event before
resuming control, such as a certain amount of simulation time passing or the
toggling of a signal. This approach is known as \textit{cosimulation}.

To verify the HCCStar's digital logic, Python driver and monitor classes
were created to control inputs and receive outputs using this approach. A Python
model of the ABCStar was written that would respond to readout commands and produce
properly-formatted data packets. Tests were written to validate all features
and probe edge cases. Code coverage tools in the Cadence simulator were
used to ensure all major features were probed by the tests.

One important subset of these tests involved checking the radiation tolerance
of the digital logic, by simulating single-event upsets (where ionizing radiation causes
a flip-flop to toggle) inside the design. The cosimulation approach made it
easy for Python tests to change the state of a flip-flop, then yield control
to the simulator and check to see if any behavioral problems develop. These
tests were particularly important due to the high dosage of radiation the ASICs
will be exposed to over the lifetime of the HL-LHC.

\begin{figure}[!htb]
    \centering
    \begin{subfigure}[t]{0.4\textwidth}
        \centering
        \includegraphics[width=\textwidth]{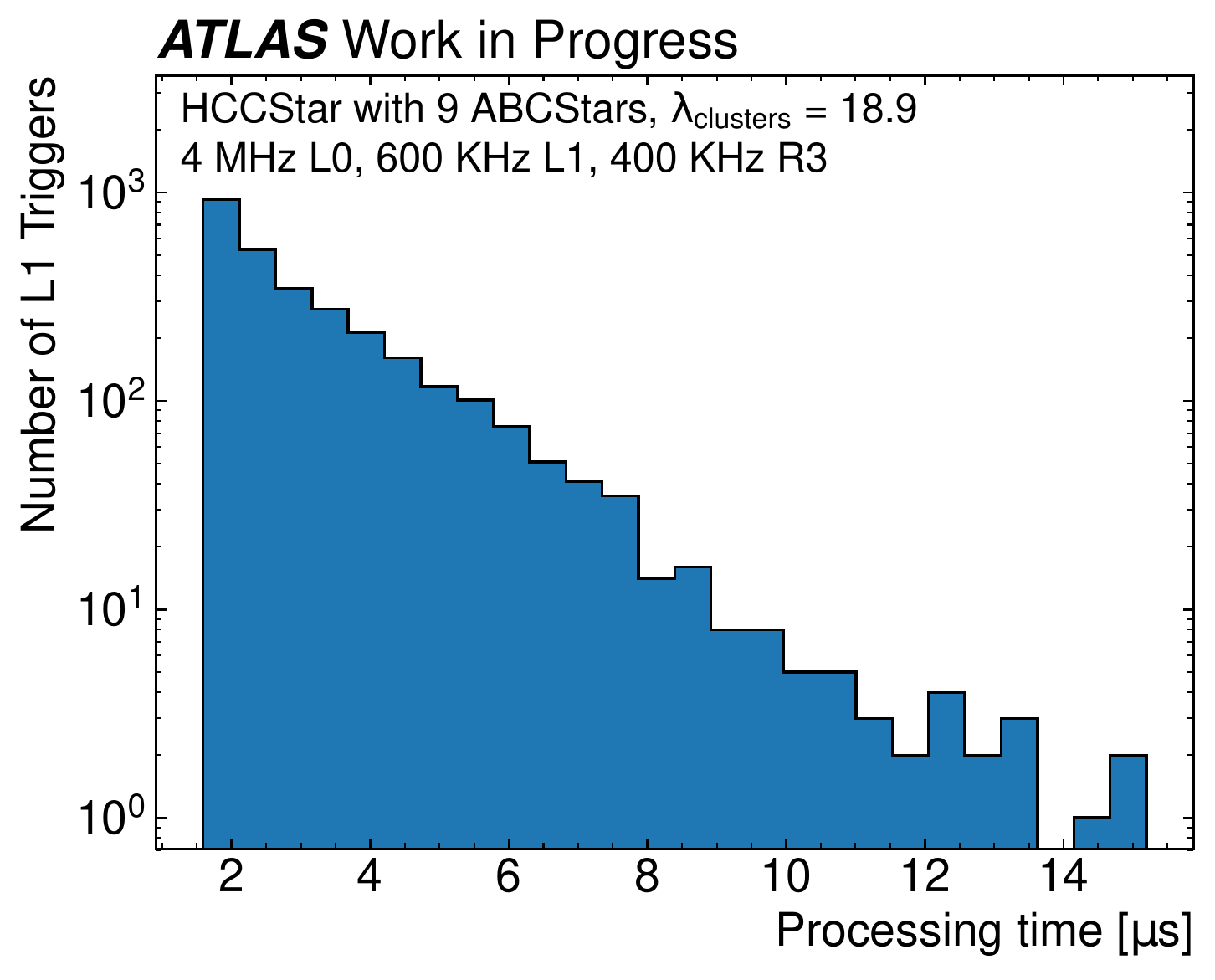}
        \caption{Total time to process L1 triggers}
    \end{subfigure}
    ~
    \begin{subfigure}[t]{0.4\textwidth}
        \centering
        \includegraphics[width=\textwidth]{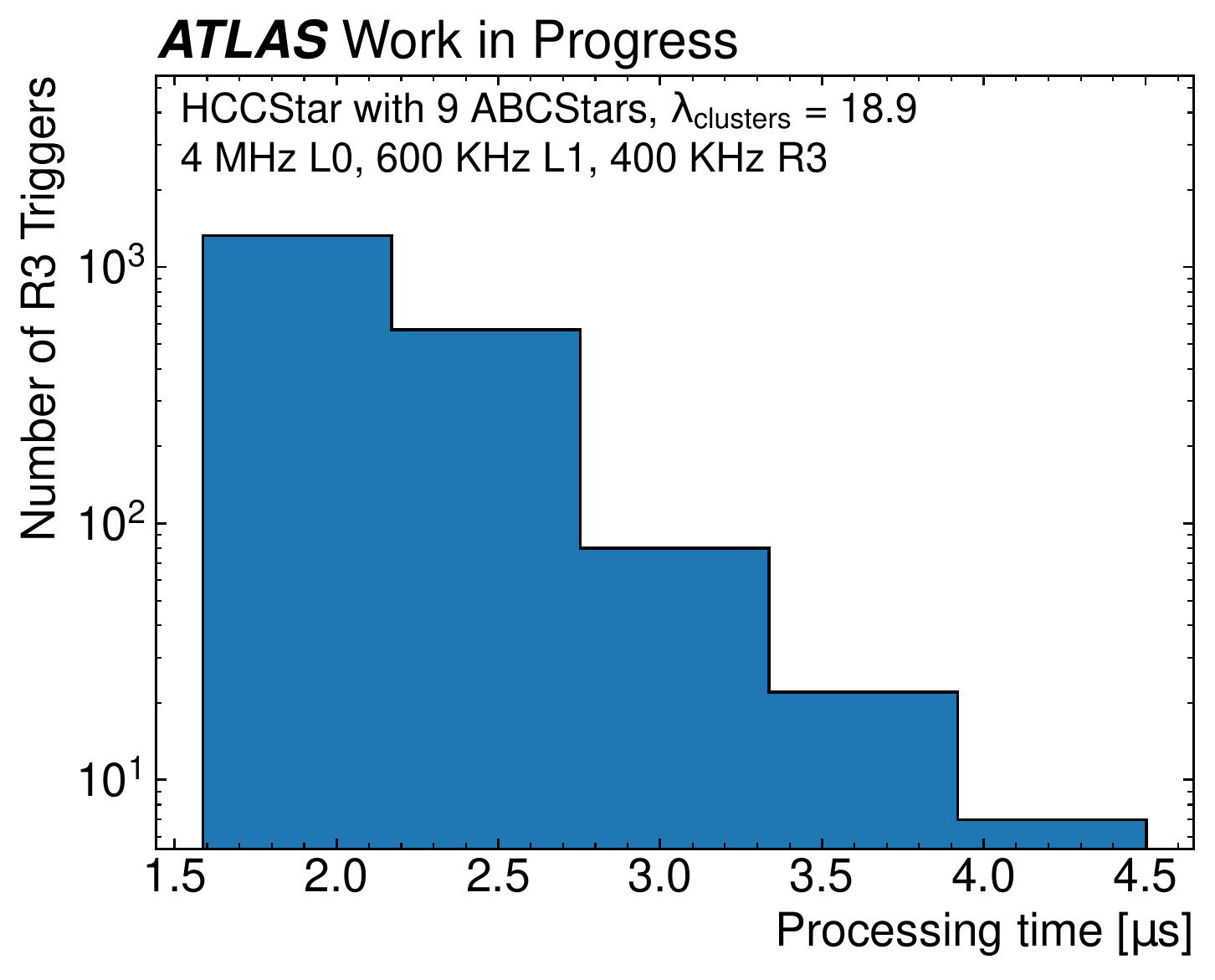}
        \caption{Total time to process R3 triggers}
    \end{subfigure}
    \caption{Example monitoring plot produced by the HCCStar testbench for
    a hybrid with nine ABCStars. This configuration shows the performance
    for the highest-occupancy modules in the strip end caps in a multi-level
    trigger scenario.}
    \label{fig:monitoring}
\end{figure}

The same testbench was also used to run combined simulations
of both the HCCStar and ABCStar designs. The Python ABCStar model was
replaced with the real ABCStar codebase and simulations of a full hybrid ran under
realistic operating conditions. Not only did this allow the interface between
the two chips to be verified, but it also enabled long, realistic simulations
of data-flow to be conducted. The use of Python made it possible to use various
scientific computing packages (PyROOT, numpy, matplotlib, etc.) from inside
the testbench, to produce monitoring plots and perform studies to check the
expected performance of the design. Figure \ref{fig:monitoring} shows example
monitoring plots in which the response time was measured for a ``worst case"
hit occupancy and readout rate.

\section{Conclusion}

The cocotb approach to digital logic verification was found to
be extremely powerful for the development of an ASIC like the HCCStar.
The use of Python made it easy for graduate students and postdocs
to contribute to the verification effort and assist the full-time engineers
responsible for the digital logic design.

During the development of the first engineering prototype, over 65 issues of
varying severity were found and fixed, highlighting the importance of thorough testing before
submission. As of this writing, the engineering prototypes have been fabricated
and are undergoing physical testing. While some revisions for the final
design are planned, the prototypes have mostly performed as expected,
indicating that the HCCStar verification using cocotb was a success.

\nocite{*}
\footnotesize{\printbibliography}

\end{document}

%% file: econfmacros.tex
\def\beq{\begin{equation}}
\def\eeq#1{\label{#1}\end{equation}}
\def\eeqn{\end{equation}}

\def\beqa{\begin{eqnarray}}
\def\eeqa#1{\label{#1}\end{eqnarray}}
\def\eeqan{\end{eqnarray}}

\let\bar=\overbar

\def\Dslash{\not{\hbox{\kern-4pt $D$}}}
\def\dslash{\not{\hbox{\kern-2pt $\del$}}}

\def\msb{{\bar{\ssstyle M \kern -1pt S}}}

